\documentclass[preprint,showpacs,preprintnumbers,amsmath,amssymb]{revtex4}

\usepackage{amsmath} 
\usepackage{amssymb}
\usepackage{epsf}
\usepackage{epsfig}
\usepackage{graphicx}
\usepackage{dcolumn}
\usepackage{bm}
\usepackage{psfrag}

\begin{document} 

\title{Nanopillar Arrays on Semiconductor Membranes as Electron Emission Amplifiers}

\author{Hua Qin$^*$, Hyun-Seok Kim, and Robert H. Blick}
\address{
Department of Electrical and Computer Engineering, 
University of Wisconsin-Madison, 
1415 Engineering Drive, Madison, WI-53706, USA
}
\date{\today}

\begin{abstract} 
A new transmission-type electron multiplier was fabricated from 
silicon-on-insulator (SOI) material 
by integrating an array of one dimensional (1D) silicon nanopillars onto  
a two dimensional (2D) silicon membrane. 
Primary electrons are injected into the nanopillar-membrane system from 
the flat surface of the membrane, while electron emission from the other side is probed by an anode.  
The secondary electron yield (SEY) from nanopillars is found to 
be about 1.8 times that of plane silicon membrane.  
This gain in electron number is slightly enhanced by the electric field applied from the anode. 
Further optimization of the dimensions of nanopillars and membrane 
and application of field emission promise an even higher gain for detector applications and 
allow for probing of electronic/mechanical excitations in nanopillar-membrane system 
excited by incident particles or radiation. 

\textnormal{$^{*}$Electronic mail: QIN1@WISC.EDU}
\end{abstract}

\maketitle

The interaction of energetic particles with matter is a fundamental process in physics. 
Many interaction pathways exist, leading to the production of a variety of secondary particles.
Arguably the most important of these processes is the so-called
secondary electron emission (SEE), which in addition to its fundamental
importance in materials properties also serves as the basis for a broad
variety of widely used practical devices (e.g. electron multipliers; cathode ray screens; 
silicon detectors)~\cite{bruining-54, wilock-60, johnson-54, kanter-61, egerton-86, stab-90, kamiya-97, lt-detectors}.
In the case of SEE stimulated by primary electrons, 
the yield of secondary electrons (SEY) is determined by the balance
between two opposing effects: on the one hand, the thicker the layer of
material interacting with the primary particles, the greater the
probability of an electron-generating event. 
However, at the same time, as the penetration depth of primary electrons increases the
ability of a secondary electron to escape from the material decreases.
This determines that SEE is an effect that mainly occurs near the surface 
and a grazing incident angle produces more secondary electrons~\cite{bruining-54, nishimura-94}. 
It has been found that a proper surface roughness can enhance
SEE while large corrugations suppress SEE. 
Recently, it has also been found that SEE can be either enhanced or
suppressed by carbon nanofibers, depending on whether nanofibers are
suspended from or attached to the underlying silicon substrate~\cite{suzuki-07}.
Unlike this random configuration of carbon nanofibers or conventional 
surface roughness produced in a sputtering process for metallic/dielectric material deposition, 
the advances in nano fabrication and material science now allow for precise
engineering of surfaces to optimize SEE and 
understanding new physics of electron impact on nano objects. 
Here we demonstrate a nanostructured material consisting of an array of one dimensional (1D) 
silicon nanopillars fabricated on the surface of a two dimensional (2D) layer of crystalline silicon. 
It is shown that this juxtaposition of structures of different
dimensionalities results in an enhanced SEE response. 
The choice of a thin membrane allows for a separation of primary and secondary electrons, i.e., 
a transmission-type electron generation. 
Naturally, this scheme is easy to extend to other materials with excellent
electron emission properties, such as diamond nanopillars, 
aligned carbon nanotubes and zinc oxide nanowires. 
This ability to alter fundamental material properties by manipulation of device geometry 
at the nanoscale level opens new opportunities for 
exploring electronic and mechanical excitations in nano structures 
and new designs of novel materials and devices.

For the purpose of this experiment we fabricated several membranes from
{\it n}-type silicon-on-insulator (SOI) wafers, as schematically shown in Fig.~1(a). 
The starting SOI material consists of a 3-micron thin layer of silicon on 
an insulating layer of silicon dioxide ($1.1~\mathrm{\mu m}$).
The substrate is of {\it n}-type silicon with a thickness of $725~\mathrm{\mu m}$.
The resistivity of the SOI is of the order of $12~\mathrm{\Omega\cdot cm}$.
Both the SOI and the silicon substrate have a crystal orientation of (100). 
After the SOI was thinned down to $2.9~\mathrm{\mu m}$ and 
thus a $250~\mathrm{nm}$ layer of silicon dioxide was formed on top by thermal oxidation, 
the whole wafer was then capped with a thin layer of silicon
nitride ($\sim 400~\mathrm{nm}$) by using low pressure chemical vapor deposition (LPCVD). 
Being chemically resistive to potassium hydroxide (KOH) solution, 
the silicon nitride coating provides an etch mask in an anisotropic etching of silicon 
to form thin silicon membranes. 
The final membranes of square shape have a side length of $35~\mathrm{\mu m}$.
On each device 16 such identical membranes were fabricated into four $2 \times 2$ arrays.
 A scanning electron micrograph of four such membranes is shown in Fig.~1(b). 
On each membrane, an array ($\approx 17,600$) of round nanopillars was fabricated 
from the membrane host by electron-beam lithography (EBL), 
gold deposition and a successive reactive-ion etching (RIE). 
Finally, the gold mask was removed in a wet chemical etch step, 
leaving clean silicon nanopillars on the membranes. 
Each pillar has a diameter of $80~\mathrm{nm}$ and a height of $300~\mathrm{nm}$. 
Close-ups of nanopillar arrays with a pitch of $200~\mathrm{nm}$ are shown in Fig. 1(c) and (d). 
In Fig. 1(e), the SEM graph of a cleaved membrane reveals the overall architecture of one-dimensional
nanopillars placed on the two-dimensional membrane. 
Also indicated in Fig. 1(b) is that the nanopillars are patterned in a frame marked
$\mathrm{\Delta}$ around the center piece of the plain membrane marked M. 
This allows to discriminate electron transmission through the membrane alone (M), 
the nanopillar-membrane system ($\mathrm{\Delta}$), 
and through the bulk material (B) which includes two extra layers of dielectrics.
The thickness of membrane (M) is about $1.6~\mathrm{\mu m}$. 

The experimental setup we used is also shown schematically in Fig.~1(a): 
the device is mounted in a scanning electron microscope (SEM) which provides a vacuum environment
($p \sim 10^{-6}~\mathrm{mbar}$) and most importantly a controllable electron beam (e-beam). 
The e-beam is scanned over the backside of the membrane to 
inject electrons in the energy range of $E_p = 1-30~\mathrm{keV}$. 
The membrane is connected to an electron reservoir at ground potential. 
A large anode is placed above the nanopillars, providing an extraction or retarding
voltage for electrons emitted from the membrane and nanopillars.
Most importantly, the anode is designed as a Faraday cup such that the
efficiency of collecting electrons approaches 100\%. 
By controlling the anode voltage ($V_a$) while monitoring the anode current ($I_a$),
secondary electron emission ($E \lesssim 50~\mathrm{eV}$) can be
differentiated from electrons transmitted through the membrane
($E \leq E_p$)~\cite{bruining-54} or from field emitted electrons ~\cite{fe}.
This provides a simple method to analyze the energy distribution of
emitted electrons and allows for identifying the effect of nanopillars on electron emission. 
This experimental setup is similar to a scanning transmission electron 
microscope (STEM)~\cite{crewe-70, browning-93}. 
However, the aim here is not to obtain an atomic resolution which requires an ultra-thin membrane.
The experimental results shown below will demonstrate that electron emission is enhanced by introducing
nanopillars on the exit side of a thin membrane.

Fig.~1(f) shows a Monte-Carlo simulation revealing 
the spatial distribution of primary electrons (colorized dots)
entering from below and the SEE (gray scale in red color) in a nanopillar-membrane device. 
In the simulation, the real dimensions were used for the nanopillar. 
The membrane thickness was chosen to be same as the height of the nanopillar 
to reduce the simulation time. The electron energy was set at $30~\mathrm{keV}$. 
In reality, the membrane could be made even thinner and behaves as a 2D system. 
As will be shown bellow, in our nano engineered nanopillar-membrane device,
it is precisely the electron-solid interaction within the nanopillars that 
enhances the overall electron generation. 
In other words, the surface increase of the 2D-membrane by 1D-nanopillars enhances 
SEE to a degree where the membrane amplifies the incoming number of 
electrons more effectively than a 3D system. 
Thus adding the dimensions 2D+1D as for the nanopillar-membrane system leads 
to a behavior different from a 3D bulk system.

Fig.~2(a) shows a color-scale map of the anode 
current (normalized by the incident beam current $I_b$)
as a function of the position of the e-beam scanning over the back side of four membranes. 
We can directly compare this map with the SEM image shown in Fig.~1(b). 
We find that the anode signal provides a high contrast in membrane thickness and 
shows a clear enhancement of electron emission 
in the area of nanopillars ($\mathrm{\Delta}$ compared to M). 
The plot in Fig.~2(b) represents a line scan taken from the
corresponding color-scale map. Obviously, one can directly follow
transitions between non-membrane area (B), membrane (M) and
membrane with nanopillars ($\mathrm{\Delta}$).

The origin of enhanced SEE from the nanopillars is further explored 
by altering the anode voltage. 
Since the anode with a negative potential will keep electrons with energy
below $e|V_a|$ from reaching the anode, it thus provides a method to analyze the
energy of emitted electrons by sweeping the anode voltage. 
The $I_a-V_a$ characteristics in Fig.~3(a) were measured for the three distinct areas
(B, M and $\mathrm{\Delta}$) for comparison. 
The anode voltage was swept from $-200~\mathrm{V}$ to $+200~\mathrm{V}$.
Constant levels of anode current are observed when $V_a<-150~\mathrm{V}$.
These levels reflect the contribution from those electrons transmitted
through the nanopillar-membrane system where the electrons' energy is
only slightly attenuated ($E \leq E_p=30~\mathrm{keV}$).
Upon further increasing the anode voltage up to $+30~\mathrm{V}$,
a continuous rise in the anode current due to SEE is found. 
Above $V_a= +30~\mathrm{V}$, most transmitted primary and secondary electrons 
are collected by the anode and the anode current reaches a saturation value. 
In Fig.~3(a), the black curve shows the electron emission 
through the non-membrane area (B), which is suppressed in reverse bias to 36\% and
increased in the forward direction to about 83\%. 
The increase of 47\% is the contribution from secondary electrons. 
Turning now to the signals from the membrane (M) and 
the nanopillar-membrane system ($\mathrm{\Delta}$),
we can see the direct transmission of the primary electrons is
increased by about 12\%, where this increase relates to 
the thinness of the membrane comparing to the unprocessed multi layers (B). 
However, the contribution from SEE is increased to 
57\% for area M and 67\% for area $\mathrm{\Delta}$.
Because of the increase in SEE, the total emission current becomes
greater than the incident current, i.e., a gain is achieved. 

We found that in contrast to the intuitive assumption {--} 
that is the thinner membrane the higher the transmission should be {--} 
a membrane with nanopillars shows an even more enhanced signal. 
As depicted in the inset of Fig.~3(a), the derivative of the
$\mathrm{\Delta}$-trace with respect to the anode voltage represents the energy distribution of
the secondary electrons. We further examined the effects of nanopillars
on electron emission by scanning the e-beam ($30~\mathrm{keV}$) 
across the nanopillar frame at $V_a = \pm 200~\mathrm{V}$. 
Electron emissions from areas B, M and $\mathrm{\Delta}$ are compared directly in Fig.~3(b). 
A remarkable influence of the nanopillars ($\mathrm{\Delta}$-peaks)
is found. Under a forward anode bias $V_a = +200~\mathrm{V}$, 
same as that in Fig.~2(b), an enhancement of SEE by the nanopillars is clearly observed. 
Under reverse anode bias $V_a = -200~\mathrm{V}$
transmission of primary electrons is slightly suppressed by the nanopillars, 
which is also seen in Fig. 3(a) (see the arrows). This is
a clear indication that the nanopillars absorb high-energy primary
electrons and generate more low-energy secondary electrons than the
2D membrane alone. 

This effect also suggests that in order to 
obtain an optimal SEY the ratio of membrane thickness to
nanopillar height and the aspect ratio of nanopillars have to be carefully tuned. 
Comparing to curve B in Fig. 3(a),  it has to be noted that curve M has a stronger 
dependence on positive anode potential. 
This is directly related to the fact that the electric field in the recessed membrane
area is retarded (see Fig.~1(a)). Furthermore, the even stronger dependence on anode
potential found in area $\mathrm{\Delta}$ stems from the suppression of the electric field 
on the nanopillar sidewall by neighboring nanopillars. 
This suggests that the SEE from a patterned/rough surface could be optimized 
by an electric field applied so that the reentrance of secondary electrons into neighboring
nanopillars is avoided. Furthermore, it is of great interest to
explore electron emission from nanopillars at even higher electrical fields 
where field emission can kick-in and help removing electrons from the nanopillars. 

The above results were obtained for the incident energy of $30~\mathrm{keV}$. 
The detailed dependence of electron emission on the incident energy is shown in Fig.~4. 
Again emission signals from areas M and $\mathrm{\Delta}$ are compared for $V_a = +200~\mathrm{V}$. 
The threshold energy for electrons to 'penetrate' the nanopillar-membrane
system is about $12.5~\mathrm{keV}$. 
There is no observable shift in the threshold energy comparing areas M and $\mathrm{\Delta}$.
However, nanopillars significantly increase the emission signal {--} that is the yield 
$\gamma=\gamma(\gamma_m, \gamma_p)$  of emitted electrons, 
where $\gamma_m$ and $\gamma_p$ are the yields of SEE for membrane and nanopillars, 
respectively~\cite{sey}.
The solid line shown in Fig. 4 is a Monte Carlo approximation to the SEE from thin membranes, 
based on the Bethe model of energy loss and 
a parametric model of SEE~\cite{bruining-54, bethe-33, joy-95}. Above $30~\mathrm{keV}$, 
which is the maximal energy available in our SEM, 
a saturation of the anode current levels is expected.

Above the threshold energy of $12.5~\mathrm{keV}$, 
an enhancement of 180\% by the nanopillars is obtained as
compared to the membrane. The cause for this enhancement obviously is
the altered surface morphology due to the nanopillars, 
which increases the effective surface area and the effective incident angle for
electrons (see Monte-Carlo simulation in Fig. 1(f)). 
It has to be noted that the thickness of current membranes is about 1.6 microns, 
which is much larger than the penetration depth of $30~\mathrm{keV}$ electrons. 
A thinner membrane allows more primary electrons to reach
the nanopillars and produce more secondary electrons. 
In the frame of this interpretation, the normalized anode current, 
defined as the total yield $\gamma = I_a/I_b$, can be expressed as 
$\gamma= \beta \gamma_p +(1-\beta) \gamma_m$,
where $\beta$ is the coverage of the membrane surface by nanopillars. 
For an area $\mathrm{\Delta}$ on this particular device, 
we have $\beta=\pi d^2/4 L^2 \approx 0.13$, where $d$ is the diameter of a nanopillar, 
and $L$ is the pitch distance. Consequently, a higher SEE can be
achieved by decreasing the pitch distance between nanopillars. 
By reducing the pitch distance from
$200~\mathrm{nm}$ to $150~\mathrm{nm}$, 
while maintaining the pillar's dimension, $\beta$ can be doubled. 
In the current device, the thickness of the membrane is much larger than the 
mean-free-path of the incident electrons meaning a large number of incident 
electrons are slowed down by scattering before they enter into nanopillars. 
Longer nanopillars will substantially increase both the generation and emission of SEs. 
A larger diameter for nanopillars increases the emission area, 
however, it also decreases the possibility for secondary electrons to escape from nanopillars. 
Hence, there is an optimized diameter corresponding to the energy of incident electrons. 
When a strong electric field is applied to prevent emitted electrons from
being absorbed by neighboring nanopillars, 
longer nanopillars (but no necessarily longer than the penetration depth) 
can  increase the emission area and hence maximize the SEY. 
Comparing to the obtained enhancement factor of 1.8 shown in Fig.~4, 
a factor of 10 in the enhancement of SEE is expectable 
if a proper optimization of the dimensions can be achieved:  
a thinner membrane, longer nanopillars, an optimized diameter and pitch distance. 
With an even higher electric field, 
field emission of stimulated electrons will take place and 
dramatically enhance the emission current~\cite{nojeh-06, qin-inpreparation}. 
A higher yield of SEE can also be realized by choosing a material with
higher intrinsic yield of SEE, e.g., diamond~\cite{geis-98}.
Here, we emphasize that integration of nanopillars on a membrane has two obvious advantages: 
(i) they naturally provide a boost to SEE by the geometrical change of the emission surface, 
as we have seen, and 
(ii) they constitute an array of pointing emitters operating in parallel,
which has great potential for including other emission mechanisms such
as electron field emission and plasmon/phonon/photon-assisted emission.

In summary we have demonstrated that 
a nanopillar-membrane system can be engineered and optimized to maximize the SEY. 
Electron-solid interactions in the world of nano objects will 
demonstrate new effects and find applications in new-concept devices.  
Particularly in the device shown here, 
the functions of the membrane and nanopillars are separated 
in a sense that the membrane acts as a filter/window for incident particles, 
while the nanopillars are the true active elements. 
It is clear that the geometry of the nanopillars and the arrays can be freely chosen. 
One can use a host of different heterostructure materials, 
such as {\it p-n} junctions, quantum wells, etc., to integrate into the nanopillar-membrane system,     
which further enhances the functionality. 
Finally, nanopillars can be further configured as electron field emitters, 
where they serve not only as a host of particle-solid interaction,  
but also as probes of electronic/mechanical excitations in nanopillar-membrane systems 
disturbed by incident particles or radiation. 

The authors thank Michael S. Westphall and Lloyd M. Smith
for helpful discussions and comments. The authors like to
acknowledge support from Wisconsin Alumni Research Foundation (WARF), 
the National Science Foundation (MRSEC-IRG1), 
and the Air Force RSO under contract number F49620-03-1-0420.


\newpage

\begin{figure}
\caption{
Schematics of the device and the experimental setup. 
The device is a thin silicon membrane with an array of nanopillars fabricated on the top side. 
(b), (c), (d), and (e) are scanning electron micrographs. 
(b) A top view of four square membranes. 
(c) and (d) are close views of a nanopillar array.
(e) A cross section of membrane.
(f) A Monte-Carlo simulation shows the different distributions of primary
electrons (colorized dots) penetrating from beneath and secondary
electrons (red color) in a nanopillar-membrane structure (see text for details). 
} \label{fig:1}

\caption{
(a) A color scale plot of anode
current as a function of the position of the scanning e-beam. 
In the experiment, the anode voltage was $+200~\mathrm{V}$.
The incident electron energy was $30~\mathrm{keV}$.
The incident e-beam current was set at $200~\mathrm{pA}$.
(b) A line scan taken between the two arrows shown in (a).
} \label{fig:2}

\caption{
(a) The anode current signals as a function of the anode voltage were
probed for comparison when the e-beam was located in areas B, M and $\Delta$.
The incident electrons had an energy of $30~\mathrm{keV}$ and the beam current was $200~\mathrm{pA}$.
The inset displays the energy distribution of secondary electrons emitted from area $\mathrm{\Delta}$.
(b) Two line scans across a single membrane and covering areas B, M and $\mathrm{\Delta}$ were taken at
$V_a = -200~\mathrm{V}$ and $V_a = +200~\mathrm{V}$ for comparison. 
For clarity, the amplitude of the line scan taken at $V_a = -200~\mathrm{V}$ 
is multiplied by a factor of 2.
} \label{fig:3}

\caption{
The dependence of SEY on the incident electron energy is compared for areas $M$
and $\mathrm{\Delta}$. The anode voltage was $+200~\mathrm{V}$.
The solid line is a Monte Carlo simulation for the membrane. 
} \label{fig:4}

\end{figure}

\newpage

\begin{figure}[!p] 
\vspace{1 cm}
\includegraphics[width=.7\textwidth]{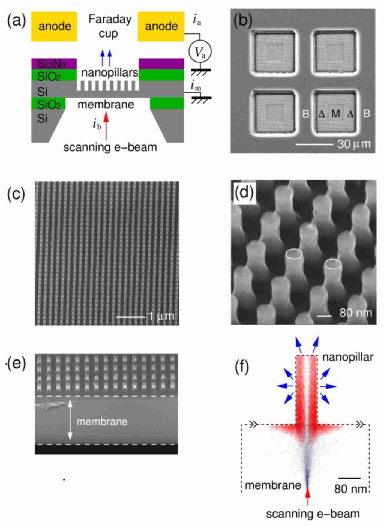}
\center{Qin {\it et al:~} Figure 1/4}
\end{figure}

\begin{figure}[!p] 
\vspace{1 cm}
\includegraphics[width=.7\textwidth]{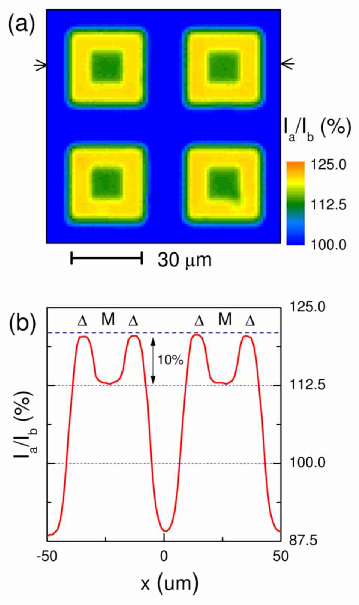}
\center{Qin {\it et al:~} Figure 2/4}
\end{figure}

\begin{figure}[!p] 
\vspace{1 cm}
\includegraphics[width=.7\textwidth]{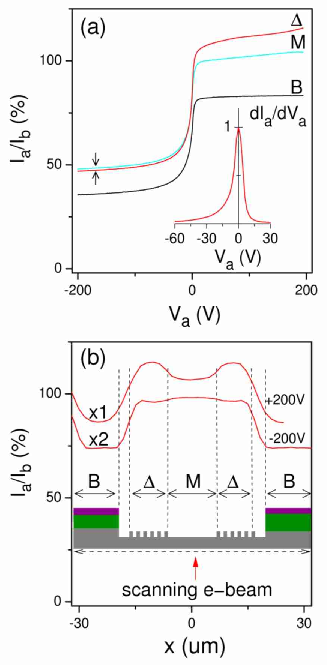}
\center{Qin {\it et al:~} Figure 3/4}
\end{figure}

\begin{figure}[!p] 
\vspace{1 cm}
\includegraphics[width=.7\textwidth]{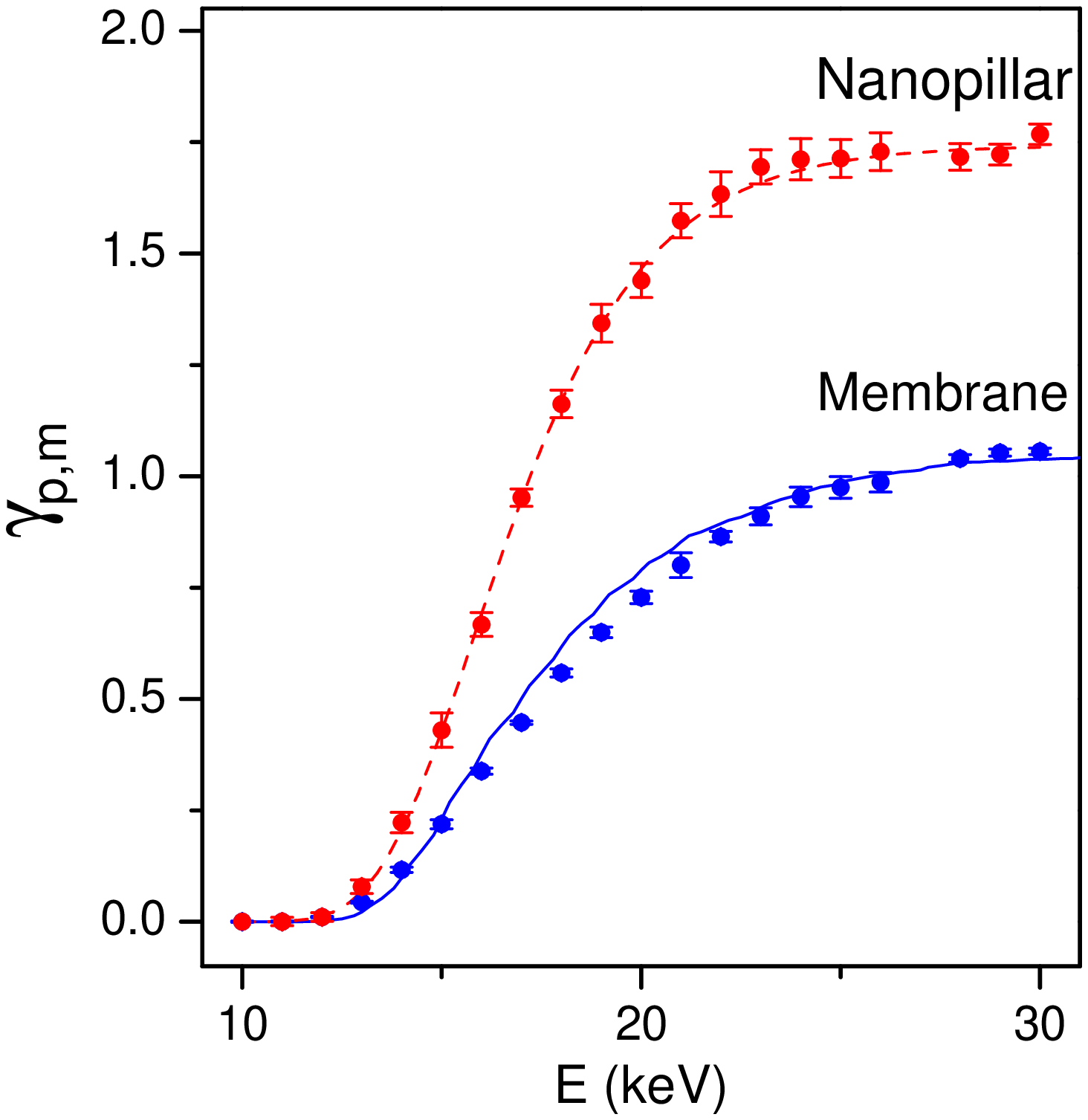}
\center{Qin {\it et al:~} Figure 4/4}
\end{figure}

\end{document}